\documentclass[prl,twocolumn,floatfix,superscriptaddress]{revtex4}

\usepackage[dvips]{graphicx}
\usepackage{epsfig}

\usepackage{verbatim}

\usepackage{times}

\usepackage{subfigure}

\usepackage{amsmath}

\usepackage{natbib}

\begin{document}

\title{Avalanche Size Scaling in Sheared Three-Dimensional 
Amorphous Solid}

\keywords{metallic glass, shear band, atomistic simulation, Mg-Cu}

\author{Nicholas P. Bailey}
\email{nbailey@ruc.dk}
\affiliation{Department of Mathematics and Physics (IMFUFA), DNRF Center 
``Glass and Time'', Roskilde University, P.O. Box 260, DK-4000 Roskilde, 
Denmark}

\author{Jakob Schi{\o}tz}
\affiliation{DNRF Center for Individual Nanoparticle Functionality, CINF, NanoDTU, Department of Physics, Technical University of Denmark, 2800 Kongens Lyngby, Denmark}

\author{ Ana\"{e}l Lema\^{i}tre}
\affiliation{Institut Navier-- LMSGC, 2 all\'ee K\'epler,77420 Champs-sur-Marne, France}

\author{ Karsten W. Jacobsen}
\affiliation{CAMP, NanoDTU, Department of Physics, Technical University of Denmark, 2800 Kongens Lyngby, Denmark}

\begin{abstract}
We have studied the statistics of plastic rearrangement events in
 a simulated amorphous solid at $T=0$. Events are characterized by the energy 
release and the ``slip volume'', the product of plastic strain and system
 volume. Their distributions for a given system size $L$ appear to be 
exponential, but a characteristic event size cannot be inferred, because the
 mean values of these quantities increase as $L^{\alpha}$ with $\alpha \sim
3/2$. In contrast to results obtained in 2D models, we do not see simply 
connected avalanches. The exponent suggests a fractal shape of the avalanches, 
which is also evidenced by the mean fractal dimension and participation ratio.
\end{abstract}

\maketitle

Athermal, or low temperature, plastic deformation of amorphous solids exhibits
intermittent stress fluctuations and shear localization, in materials as 
diverse as metallic glasses \cite{Schuh/Nieh:2003}, granular 
materials \cite{Miller/OHern/Behringer:1997}, foams \cite{Pratt/Dennin:2003} 
and glassy polymers \cite{Haward/Young:1997}. Detailed knowledge of plastic 
deformation mechanisms in amorphous solids, and their connection to macroscopic
 flow properties, however, remains elusive: While in crystals the dislocation
 provides a well defined starting point for estimates of flow stress, 
$\sigma_f$, in glasses there is no such easily characterizable defect. 
The traditional picture of deformation in amorphous solids--pioneered by 
Argon \cite{Argon79} and coworkers--is that plasticity involves 
collections of `relaxation centers' \cite{Khonik/others:1998} or 
`shear-transformation zones'(STZs) \cite{Falk/Langer:1998} which operate as 
localized centers of deformation. This picture is supported by simulations of
 deformation in amorphous metals \cite{Maeda/Takeuchi:1981, 
Srolovitz/Vitek/Egami:1982,
Falk/Langer:1998, Malandro/Lacks:1999, Lund/Schuh:2003a, Lund/Schuh:2003b} and 
observations of localized 
events \cite{Srolovitz/Vitek/Egami:1982, Zink/others:2006}.

Mean-field theories of plasticity \cite{Khonik/others:1998,Falk/Langer:1998,
lemaitre02b} rely on this viewpoint, with the additional assumption that STZs 
operate somewhat independently. But the detailed nature of correlations between
 shear transformations is a subtle issue. Elementary shear transformations 
should give rise to long-range elastic displacement fields, in analogy with the
 transformation of elliptic Eshelby inclusions, in particular, a $1/r^3$ stress 
field due to a compact source. Models incorporating such interactions
\cite{BulatovArgon,BaretVR02} exhibit localization of deformation in patterns 
reminiscent of shear bands \cite{Langer01}, suggesting that long-range elastic 
interactions may play an important role in the plastic response. The occurrence
 of shear bands in metallic glasses has been a major obstacle in the'
 development of these materials for engineering 
applications \cite{Johnson:1999}.

Even in carefully prepared samples (free of fracture-producing flaws), 
experimental observation of plastic deformation is often hindered by 
localization. In numerical simulations, however, it is possible to preserve 
translation-invariance and access statistical properties of plasticity in 
steady state. Athermal, quasi-static deformation allows further simplification
of the underlying dynamics: Lacks has shown that in potential energy 
landscape (PEL), plastic deformation involves the destabilization of local 
minima along a single zero mode
\cite{Malandro/Lacks:1998,Malandro/Lacks:1999, Lacks:2001}.
The PEL point of view brought hope that elementary shear transformations
could be identified with elementary transitions between minima in the PEL, as 
often held by STZ theories. This notion has been challenged, however, in recent
simulations in two and three dimensions (2D and 3D). These show that individual
 plastic events present multiple substructures, which are
more compact and localized \cite{Maloney/Lemaitre,Demkowicz/Argon}. Maloney and
Lema\^{i}tre \cite{Maloney/Lemaitre} found that when visualized according to 
active atoms, plastic events--transitions between local minima--tended to be 
localized in one dimension but spread out along the other. This behavior led to
 an apparent scaling where the energy released in an event scaled as $L$, the 
linear system size. 

It is essential to know whether the 2D results of Ref.~\cite{Maloney/Lemaitre}
should transfer to 3D. This is not obvious because the 
correlations that lead to an event taking place over an extended region
may depend crucially on the power-law dependence of the elastic Green's
function which is weaker in 3D. In this Letter, we report 
3D simulations of a realistic model metallic glass, 
undergoing athermal quasi-static shear deformation.
The main novelties of our work are: (i) the dimensionality; 
(ii) our use of realistic interactions potentials; 
(iii) our use of fractal analysis to characterize the geometry of avalanches 
in 3D. Our main results are (1) we 
observe a scaling of event sizes with exponent close to 3/2; (2)
visualization of typical large events indicates that the avalanches are no
longer ``simply connected'', partially localized avalanches, but rather are
spread throughout the simulation box, with an apparent fractal shape, with
mean fractal dimension close to 3/2. We note that this scaling behavior we 
differs from that observed in studies of crackling noise
 in magnets, dislocation avalanches in single-crystal 
plasticity \cite{Weiss/Marsan:2003} or other systems that are characterized by
 critical behavior which leads to power law distributions of event sizes 
\cite{Sethna/Dahmen/Myers:2001}. For such systems finite-size effects only 
influence the large-avalanche tail of the power law distribution. In our case
 the delocalized nature of the avalanches leads to a situation where the whole
distribution scales with system size.

The simulated material is Mg$_{0.85}$Cu$_{0.15}$, which is the optimal 
glass-forming composition for the Mg-Cu system 
\cite{Sommer/Bucher/Predal:1980}. This system is interesting because the 
addition of a small amount of Y makes it a bulk metallic glass with high 
strength and low weight \cite{Inoue/others:1991}. The interatomic potential is 
the effective medium theory \cite{Jacobsen/Stoltze/Norskov:1996}, fitted to 
properties of the pure elements and intermetallic compounds obtained from 
experiment and density functional theory calculations. The configurations were
 created by cooling from a liquid state above the melting temperature down to 
T=0, using constant temperature and pressure molecular dynamics. Details of the
 potential and of the cooling process may be found in
\cite{Bailey/Schiotz/Jacobsen:2004a}; the cooling rate for the systems studied 
here was about $10^{11}$Ks$^{-1}$. Periodic boundary conditions were used both 
in cooling and in the deformation simulations described below. Five system 
sizes were studied, containing 864, 2048, 4000, 8788 and 16384 atoms, with
 $L=26$\AA, 35\AA, 44\AA, 57\AA\ and 70\AA, respectively. Ten independent 
configurations were produced for all sizes, except only one 16384-atom system. 

The systems were deformed in pure shear by a variation of the standard 
procedure of straining the entire system homogeneously in small increments 
followed by energy-minimization. In the so-called quasi-static limit, the 
system continuously follows deformation-induced changes in local 
minima \cite{Malandro/Lacks:1998, Malandro/Lacks:1999}; this protocol is meant 
to capture the asymptotic trajectory in this limit. During homogeneous strains,
as well as relaxing the atomic positions, components of strain apart from the 
one being controlled were also relaxed. In particular this means that the 
hydrostatic pressure was always zero. To observe scaling it is important to be 
close to the quasi-static limit and therefore to have a strict tolerance 
for minimization \cite{Maloney/Lemaitre}; we used 10$^{-6}$eV/\AA\ for 
the maximum force and 10$^{-6}$eV/\AA$^3=0.16$ MPa for the maximum (relaxable) 
stress. The homogeneous strain was applied by multiplying the simulation box 
vectors by a shear strain matrix at each step, with a strain size of 
$\Delta\epsilon=0.0005$. Here $\epsilon$ is an off-diagonal component of the 
strain, not the engineering strain. The total strain at any point in the 
deformation history is the number of steps times 0.0005. The total amount of
 deformation imposed was $\sim$100\%. The use of periodic boundaries and the
small sizes prevent any kind of macroscopic localization being observed even at
such large strains.


\begin{figure}
\epsfig{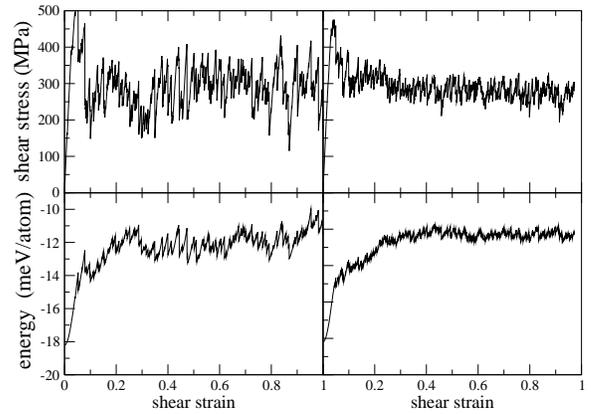}
\caption{\label{stressStrain}Stress and energy versus strain for 2048- 
(left) and 16384-atom (right) systems.}
\end{figure}


Examples of stress-strain curves are shown in Fig.~\ref{stressStrain}. The
behavior is very similar to that observed in other simulations in 2D and
3D \cite{Maloney/Lemaitre,Malandro/Lacks:1998, Malandro/Lacks:1999}: an initial
 linear elastic regime, followed by the onset of plasticity manifested as 
abrupt drops. After about 30\% strain a steady state has been reached (note in 
particular the energy of the 16384-atom system). The stress averaged over the 
steady-state part of the curve, $\sigma_f$, is about 280 MPa, independent of 
$L$.

The stress- and energy-drops define distinct plastic ``events''. These do not 
necessarily correspond to the elementary units of plastic deformation, as 
mentioned above; a complete event is an avalanche of sub-events. Defining 
subevents in a useful way is problematic in practice, hence in order to stick 
with meaningful and well-defined quantities, we study complete events.


%

We have analyzed the events in the stress and energy curves by assigning to 
each event two quantities. The energy and stress drops are defined relative to 
the values they would be expected to have given continued elastic behavior. 
Thus we have $\Delta\sigma_{drop} = \sigma_i + 2\mu\Delta\epsilon - 
\sigma_{i+1}$, where $\mu$ is the shear modulus (determined as half the slope
 of the stress-strain curve), and $\Delta E_{drop} = E_i + 
V_i\sigma_i\Delta\epsilon - E_{i+1}$, where $V_i$ is the system volume. $\Delta
 E_{drop}$ ($\Delta\sigma_{drop}$) is positive if there is a drop in stress 
(energy). For some very small stress drops, the apparent energy 
drop is negative, due to finite resolution implied by a finite 
$\Delta\epsilon$. We count events with $\Delta\sigma_{drop}$ greater 
than a cutoff $\Delta\sigma_{min}$ and $\Delta E_{drop} > 0$. Our scaling 
analysis is based on the distributions of $\Delta E_{drop}$
and a quantity proportional to $\Delta\sigma_{drop}$, that we call the 
slip volume, $V_{slip} \equiv V \Delta\sigma_{drop}/2\mu = V 
(\Delta\epsilon-(\sigma_{i+1}-\sigma_i)/2\mu)= 
V(\Delta\epsilon-\Delta\epsilon_{el}) = V\Delta\epsilon_{pl}$, where 
$\Delta\epsilon_{el}$ ($\Delta\epsilon_{pl}$) is elastic (plastic) strain. The
 significance of $V_{slip}$ can be understood by supposing 
first that the plastic slip associated with an event is confined to a 
localized region of space, whose size had a narrow distribution independent 
of $L$. If the slip is characterized as a displacement $d$ over an area $A$,
 then  $\epsilon_{pl} = Ad/V$ and $V_{slip} = Ad$. If an event involved $m$ 
such elementary shear transformations the resulting $V_{slip}$ would be 
$\sim mAd$, and thus a measure of the number of elementary transformations 
which contribute to the macroscopic stress relaxation.
The cutoff $\Delta\sigma_{min}$ is chosen so that the minimum 
$V_{slip}$ is independent of $L$ and equal to 5\AA$^3$.


\begin{figure}
\epsfig{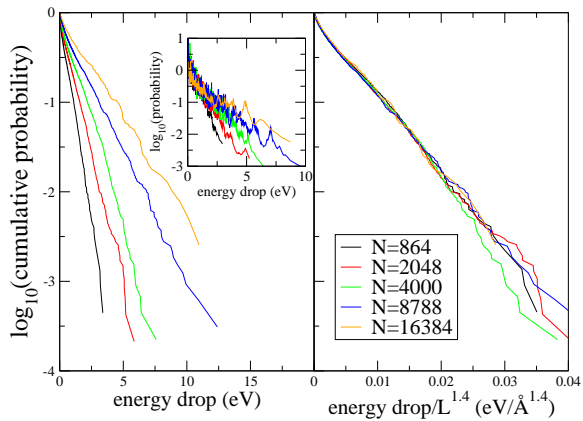}
\caption{\label{energyDropsCollapse} (Color online) Left, cumulative
 probability distribution of $\Delta E_{drop}$ in plastic events for different 
system sizes. Right, cumulative probability distribution of
 $\Delta E_{drop}/L^{1.4}$. Inset, probability distribution of
 $\Delta E_{drop}$.}
\end{figure}

\begin{figure}
\epsfig{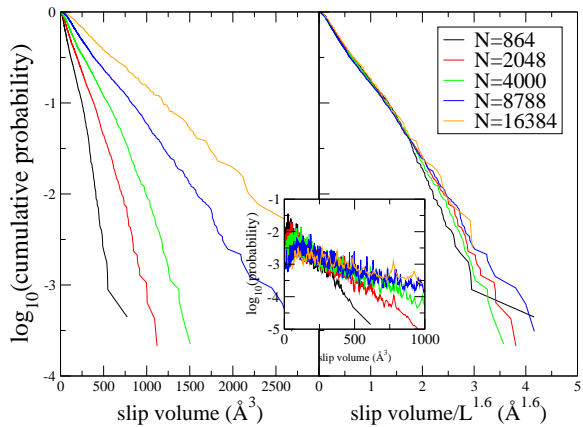}
\caption{\label{slipVolumesCollapse} (Color online) Left, cumulative
 probability distribution of $\Delta V_{slip}$ in plastic events for different 
system sizes. Right, cumulative probability distribution of 
$\Delta V_{slip}/L^{1.6}$. Inset, probability distribution of $\Delta V_{slip}$.}
\end{figure}

In Figs.~\ref{energyDropsCollapse} and~\ref{slipVolumesCollapse} are shown 
cumulative distributions $C(x)$ of $\Delta E$ and $V_{slip}$, where 
$C(x) = \int_x^\infty P(x') dx'$, and $P(x)$ is the probability distribution. 
The advantage of using $C(x)$ is that it yields a much smoother curve, while 
no information is lost through  binning. Furthermore, for power-law or 
exponential behavior of $P(x)$ at large $x$, $C(x)$ maintains this behavior
(with the exponent changing by one in the case of a power-law). The insets show
$P(x)$, which involves a binning procedure and result in noisy curves. The 
$C(x)$ curves are almost linear (with perhaps some downward curvature) in these
semi-log plots, suggesting that the distributions are roughly exponential. The
 inverse-slopes, and hence the mean values, however, systematically increase 
with $L$, so they cannot be associated with a characteristic event size
independent of $L$. If these quantities are scaled by $L^{1.4}$ and $L^{1.6}$, 
respectively, the C(x) curves collapse quite well onto master 
curves, as shown in the right panels of Figs.~\ref{energyDropsCollapse} 
and~\ref{slipVolumesCollapse}. Changing the exponent by 0.1 produces a
slightly worse collapse in both cases. The means of $\Delta E$ and $V_{slip}$ 
scale as $L^{1.43\pm0.03}$ and $L^{1.63\pm0.04}$, respectively; these 
exponents are consistent with the scaling collapses.


Ref.~\cite{Maloney/Lemaitre} interpreted the linear scaling with $L$ in terms 
of the geometrical structure of the events: they tended to be extended in one 
dimension, in the form of slip lines passing through the simulation cell.
Extrapolating their results to 3D, one might expect planar events, scaling as 
$L^2$. This is excluded by our results. If the connection between the observed
 scaling of event distributions and the geometry of events is to be trusted, we
 can tentatively interpret the $L^{3/2}$ scaling as reflecting a fractal 
geometry of the avalanches, somewhere between string-like and planar. Of
 course, the scaling of the form $L^{3/2} = V^{1/2}$ is very reminiscent of a 
central limit theorem---the variance in the extensive quantities $E$ and 
$V\sigma$ should go like $V$ (or $N$), and $\overline{\Delta E}$ and 
$\overline{\Delta\sigma}$ correspond to the square root, the standard 
deviation. Note that the results of Ref.~\cite{Maloney/Lemaitre} are also 
consistent with this interpretation, since in two dimensions $L$ is the square
root of the system size. A central limit interpretation of this scaling would
suggest that contributions from different parts of the system somehow add up in
 an uncorrelated way. The clear change in event size distribution with system 
size indicates that the events are spatially delocalized and in this way the 
resulting energy and stress drops certainly result from contributions from
 different parts of space. But  within such a ``random-noise'' interpretation
 with uncorrelated contributions from different parts of space one would expect
that the atoms participating in the event would be more or less equally
distributed over space. We  shall see now that this is not the case.

\begin{figure}
\epsfig{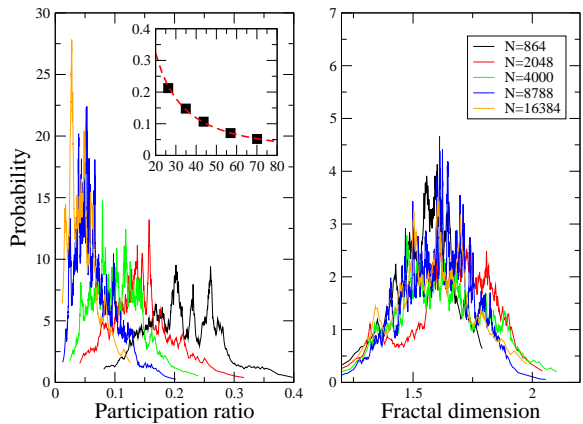}
\caption{\label{partRatioFracDim} (Color online) Left, distribution of 
participation ratios $P$ for different system sizes $L$. Inset, the 
mean $P$ versus $L$ together with a power law fit with exponent -1.44. Right,
distribution of apparent
fractal dimension of participating atoms.}
\end{figure}

To study the geometrical structure of events, we need to define a measure of 
which atoms take part in an event. We choose to consider the atom displacements
 that take place during minimization $d_i \equiv |\Delta \vec r_i|$. Rather 
than impose an arbitrary cut-off to identify participating atoms, we use the 
participation ratio $P = (\sum_i d_i^2)^2  / (N \sum_i d_i^4)$, where $N$ is 
the number of atoms. $P=1$ for an event where all the $d_i$ are equal and
 $1/N$ for one where a single atom moves; thus it is the effective fraction of 
participating atoms. In the left panel of Fig.~\ref{partRatioFracDim} we show 
distributions of $P$ for different $L$, as well as the means. The mean value 
of $P$ is well fit by a power-law with exponent -1.44$\pm$0.03. To compare with
the scaling of the extensive quantities $\Delta E$ and $V_{slip}$, we should 
consider $V P=L^3 P$, or add three to the exponent. This gives 1.56, 
close to the $V_{slip}$ exponent, 
implying a similar scaling for two measures of events which may both be 
considered ``geometrical'' in a sense---we saw above that $V_{slip}$ is 
related to the amount of plastic strain at the boundaries that an event 
causes.

Given $P$ for an event, we define the set of set of participating atoms as
those whose $d_i$ is in the top $P$ of the population. This defines the 
``mobile'' atoms without an arbitrary cut-off. We then define the fractal 
dimension $D_F$ of the set in the usual box-counting way: Taking a box which 
contains all of the atoms, we divide the box by increasing powers of two 
(in all directions) and count how many boxes $N_b$ for a given divisor $d$ are 
required to contain all atoms in the subset. $D_F$ is 
determined as the best-fit slope on a plot of $log(N_b)$ versus $log(d)$, 
assuming the data lie on a straight line. Typically four points are
available for the fit. The data generally exhibit a noticeable downward 
curvature, so the interpretation of a fractal geometry should not
be taken too literally. Even so, it is interesting that the average $D_F$ 
computed this way indeed gives the value 1.6$\pm 0.05$, independent of system
size and close to the exponents determined from $V_{slip}$ and $P$; 
distributions of $D_F$ are shown in the right hand panel of 
Fig.~\ref{partRatioFracDim}. These observations support the idea that 
non-local plastic events can still be viewed as sums of similar elementary 
events, probably like elementary shear transformations. These sub-events 
organize in space along a fractal-like structure, and therefore must be 
strongly correlated. This suggests that mean-field approaches, such as the 
STZ theory\cite{Falk/Langer:1998}, may be incomplete.


In summary we have observed a clear signature of a $\sim$3/2 scaling of event
sizes with system size, and presented evidence that this can be attributed to
a fractal-like shape of the events. The range of system sizes is not large, but
the statistics are good. It is interesting to speculate whether there is a
 relation between the events seen in low temperature deformation and those in 
a supercooled liquid near the glass transition. The latter are known to exhibit
strong spatial correlations; a recent attempt to infer a fractal dimension for
 individual clusters \cite{VollmayrLee/Zippelius:2005} yielded a value of 
1.8---not close  enough to 1.5 to suggest an obvious connection, but enough
perhaps to speculate that the dynamics changes in a smooth way upon going from
 high-$T$, zero stress to zero-$T$, high stress. Finally we mention the
 interesting question of how finite strain rates and temperatures cut off the 
scaling, and whether a length scale emerges from this cutting off, which could
 be connected to, for example, the observed width of shear bands (10--20 nm).

\begin{acknowledgments}
We acknowledge useful discussions with J. P. Sethna.
This work was supported by the EU Network on bulk metallic glass composites
(MRTN-CT-2003-504692 ``Ductile BMG Composites'') and by the Danish Center for 
Scientific Computing through Grant No. HDW-1101-05. Center for viscous liquid 
dynamics ``Glass And Time'' and Center for Individual Nanoparticle 
Functionality (CINF) are sponsored by The Danish National Research Foundation.
\end{acknowledgments}


\begin{thebibliography}{29}
\expandafter\ifx\csname natexlab\endcsname\relax\def\natexlab#1{#1}\fi
\expandafter\ifx\csname bibnamefont\endcsname\relax
  \def\bibnamefont#1{#1}\fi
\expandafter\ifx\csname bibfnamefont\endcsname\relax
  \def\bibfnamefont#1{#1}\fi
\expandafter\ifx\csname citenamefont\endcsname\relax
  \def\citenamefont#1{#1}\fi
\expandafter\ifx\csname url\endcsname\relax
  \def\url#1{\texttt{#1}}\fi
\expandafter\ifx\csname urlprefix\endcsname\relax\def\urlprefix{URL }\fi
\providecommand{\bibinfo}[2]{#2}
\providecommand{\eprint}[2][]{\url{#2}}

\bibitem[{\citenamefont{Schuh and Nieh}(2003)}]{Schuh/Nieh:2003}
\bibinfo{author}{\bibfnamefont{C.~A.} \bibnamefont{Schuh}} \bibnamefont{and}
  \bibinfo{author}{\bibfnamefont{T.~G.} \bibnamefont{Nieh}},
  \bibinfo{journal}{Acta Mater.} \textbf{\bibinfo{volume}{51}},
  \bibinfo{pages}{87} (\bibinfo{year}{2003}).

\bibitem[{\citenamefont{Miller et~al.}(1997)\citenamefont{Miller, O'Hern, and
  Behringer}}]{Miller/OHern/Behringer:1997}
\bibinfo{author}{\bibfnamefont{B.}~\bibnamefont{Miller}},
  \bibinfo{author}{\bibfnamefont{C.}~\bibnamefont{O'Hern}}, \bibnamefont{and}
  \bibinfo{author}{\bibfnamefont{R.~P.} \bibnamefont{Behringer}},
  \bibinfo{journal}{Phys. Rev. Lett.} \textbf{\bibinfo{volume}{77}},
  \bibinfo{pages}{3110} (\bibinfo{year}{1997}).

\bibitem[{\citenamefont{Pratt and Dennin}(2003)}]{Pratt/Dennin:2003}
\bibinfo{author}{\bibfnamefont{E.}~\bibnamefont{Pratt}} \bibnamefont{and}
  \bibinfo{author}{\bibfnamefont{M.}~\bibnamefont{Dennin}},
  \bibinfo{journal}{Phys. Rev. E} \textbf{\bibinfo{volume}{67}},
  \bibinfo{pages}{51402} (\bibinfo{year}{2003}).

\bibitem[{\citenamefont{Haward and Young}(1997)}]{Haward/Young:1997}
\bibinfo{author}{\bibfnamefont{R.~N.} \bibnamefont{Haward}} \bibnamefont{and}
  \bibinfo{author}{\bibfnamefont{R.~J.} \bibnamefont{Young}},
  \emph{\bibinfo{title}{{The Physics of Glassy Polymers}}}
  (\bibinfo{publisher}{Chapman and Hall}, \bibinfo{year}{1997}).

\bibitem[{\citenamefont{Argon}(1979)}]{Argon79}
\bibinfo{author}{\bibfnamefont{A.~S.} \bibnamefont{Argon}},
  \bibinfo{journal}{Acta Met} \textbf{\bibinfo{volume}{27}},
  \bibinfo{pages}{47} (\bibinfo{year}{1979}).

\bibitem[{\citenamefont{Khonik et~al.}(1998)\citenamefont{Khonik, Kosilov,
  Mikhailov, and Sviridov}}]{Khonik/others:1998}
\bibinfo{author}{\bibfnamefont{V.~A.} \bibnamefont{Khonik}},
  \bibinfo{author}{\bibfnamefont{A.~T.} \bibnamefont{Kosilov}},
  \bibinfo{author}{\bibfnamefont{V.~A.} \bibnamefont{Mikhailov}},
  \bibnamefont{and} \bibinfo{author}{\bibfnamefont{V.~V.}
  \bibnamefont{Sviridov}}, \bibinfo{journal}{Acta. Mater.}
  \textbf{\bibinfo{volume}{46}}, \bibinfo{pages}{3399} (\bibinfo{year}{1998}).

\bibitem[{\citenamefont{Falk and Langer}(1998)}]{Falk/Langer:1998}
\bibinfo{author}{\bibfnamefont{M.~L.} \bibnamefont{Falk}} \bibnamefont{and}
  \bibinfo{author}{\bibfnamefont{J.~S.} \bibnamefont{Langer}},
  \bibinfo{journal}{Phys. Rev. E} \textbf{\bibinfo{volume}{57}},
  \bibinfo{pages}{7192} (\bibinfo{year}{1998}).

\bibitem[{\citenamefont{Maeda and Takeuchi}(1981)}]{Maeda/Takeuchi:1981}
\bibinfo{author}{\bibfnamefont{K.}~\bibnamefont{Maeda}} \bibnamefont{and}
  \bibinfo{author}{\bibfnamefont{S.}~\bibnamefont{Takeuchi}},
  \bibinfo{journal}{Philos. Mag. A} \textbf{\bibinfo{volume}{44}},
  \bibinfo{pages}{643} (\bibinfo{year}{1981}).

\bibitem[{\citenamefont{Srolovitz et~al.}(1982)\citenamefont{Srolovitz, Vitek,
  and Egami}}]{Srolovitz/Vitek/Egami:1982}
\bibinfo{author}{\bibfnamefont{D.}~\bibnamefont{Srolovitz}},
  \bibinfo{author}{\bibfnamefont{V.}~\bibnamefont{Vitek}}, \bibnamefont{and}
  \bibinfo{author}{\bibfnamefont{T.}~\bibnamefont{Egami}},
  \bibinfo{journal}{Acta. Metall.} \textbf{\bibinfo{volume}{31}},
  \bibinfo{pages}{335} (\bibinfo{year}{1982}).

\bibitem[{\citenamefont{Malandro and Lacks}(1999)}]{Malandro/Lacks:1999}
\bibinfo{author}{\bibfnamefont{D.~L.} \bibnamefont{Malandro}} \bibnamefont{and}
  \bibinfo{author}{\bibfnamefont{D.~J.} \bibnamefont{Lacks}},
  \bibinfo{journal}{J. Chem. Phys.} \textbf{\bibinfo{volume}{110}},
  \bibinfo{pages}{4593} (\bibinfo{year}{1999}).

\bibitem[{\citenamefont{Lund and Schuh}(2003{\natexlab{a}})}]{Lund/Schuh:2003a}
\bibinfo{author}{\bibfnamefont{A.~C.} \bibnamefont{Lund}} \bibnamefont{and}
  \bibinfo{author}{\bibfnamefont{C.~A.} \bibnamefont{Schuh}},
  \bibinfo{journal}{Nature Mater.} \textbf{\bibinfo{volume}{2}},
  \bibinfo{pages}{449} (\bibinfo{year}{2003}{\natexlab{a}}).

\bibitem[{\citenamefont{Lund and Schuh}(2003{\natexlab{b}})}]{Lund/Schuh:2003b}
\bibinfo{author}{\bibfnamefont{A.~C.} \bibnamefont{Lund}} \bibnamefont{and}
  \bibinfo{author}{\bibfnamefont{C.~A.} \bibnamefont{Schuh}},
  \bibinfo{journal}{Acta Mater.} \textbf{\bibinfo{volume}{51}},
  \bibinfo{pages}{5399} (\bibinfo{year}{2003}{\natexlab{b}}).

\bibitem[{\citenamefont{Zink et~al.}(2006)\citenamefont{Zink, Samwer, Johnson,
  and Mayr}}]{Zink/others:2006}
\bibinfo{author}{\bibfnamefont{M.}~\bibnamefont{Zink}},
  \bibinfo{author}{\bibfnamefont{K.}~\bibnamefont{Samwer}},
  \bibinfo{author}{\bibfnamefont{W.~L.} \bibnamefont{Johnson}},
  \bibnamefont{and} \bibinfo{author}{\bibfnamefont{S.~G.} \bibnamefont{Mayr}},
  \bibinfo{journal}{Phys. Rev. B} \textbf{\bibinfo{volume}{73}},
  \bibinfo{pages}{172203} (\bibinfo{year}{2006}).

\bibitem[{\citenamefont{Lema\^{\i}tre}(2002)}]{lemaitre02b}
\bibinfo{author}{\bibfnamefont{A.}~\bibnamefont{Lema\^{\i}tre}},
  \bibinfo{journal}{Phys. Rev. Lett.} \textbf{\bibinfo{volume}{89}},
  \bibinfo{pages}{195503} (\bibinfo{year}{2002}).

\bibitem[{\citenamefont{Bulatov and Argon}(1994{\natexlab{a}})}]{BulatovArgon}
\bibinfo{author}{\bibfnamefont{V.~V.} \bibnamefont{Bulatov}} \bibnamefont{and}
  \bibinfo{author}{\bibfnamefont{A.~S.} \bibnamefont{Argon}},
  \bibinfo{journal}{Model. Simul. Mater. Sci. Eng.}
  \textbf{\bibinfo{volume}{2}}, \bibinfo{pages}{167}
  (\bibinfo{year}{1994}{\natexlab{a}}), \bibinfo{author}{\bibfnamefont{V.~V.} 
\bibnamefont{Bulatov}} \bibnamefont{and} \bibinfo{author}{\bibfnamefont{A.~S.} 
\bibnamefont{Argon}}, \bibinfo{journal}{Model. Simul. Mater. Sci. Eng.}
  \textbf{\bibinfo{volume}{2}}, \bibinfo{pages}{185}, 
\bibinfo{author}{\bibfnamefont{V.~V.} \bibnamefont{Bulatov}} \bibnamefont{and}
  \bibinfo{author}{\bibfnamefont{A.~S.} \bibnamefont{Argon}},
  \bibinfo{journal}{Model. Simul. Mater. Sci. Eng.}
  \textbf{\bibinfo{volume}{2}}, \bibinfo{pages}{203}
  (\bibinfo{year}{1994}{\natexlab{c}}).

\bibitem[{\citenamefont{Baret et~al.}(2002)\citenamefont{Baret, Vandembroucq,
  and Roux}}]{BaretVR02}
\bibinfo{author}{\bibfnamefont{J.~C.} \bibnamefont{Baret}},
  \bibinfo{author}{\bibfnamefont{D.}~\bibnamefont{Vandembroucq}},
  \bibnamefont{and} \bibinfo{author}{\bibfnamefont{S.}~\bibnamefont{Roux}},
  \bibinfo{journal}{Phys. Rev. Lett.} \textbf{\bibinfo{volume}{89}},
  \bibinfo{pages}{195506} (\bibinfo{year}{2002}).

\bibitem[{\citenamefont{Langer}(2001)}]{Langer01}
\bibinfo{author}{\bibfnamefont{J.~S.} \bibnamefont{Langer}},
  \bibinfo{journal}{Phys. Rev. E} \textbf{\bibinfo{volume}{64}},
  \bibinfo{pages}{011504} (\bibinfo{year}{2001}).

\bibitem[{\citenamefont{Johnson}(1999)}]{Johnson:1999}
\bibinfo{author}{\bibfnamefont{W.~L.} \bibnamefont{Johnson}},
  \bibinfo{journal}{MRS Bulletin} \textbf{\bibinfo{volume}{24}},
  \bibinfo{pages}{42} (\bibinfo{year}{1999}).

\bibitem[{\citenamefont{Malandro and Lacks}(1998)}]{Malandro/Lacks:1998}
\bibinfo{author}{\bibfnamefont{D.~L.} \bibnamefont{Malandro}} \bibnamefont{and}
  \bibinfo{author}{\bibfnamefont{D.~J.} \bibnamefont{Lacks}},
  \bibinfo{journal}{Phys. Rev. Lett.} \textbf{\bibinfo{volume}{81}},
  \bibinfo{pages}{5576} (\bibinfo{year}{1998}).

\bibitem[{\citenamefont{Lacks}(2001)}]{Lacks:2001}
\bibinfo{author}{\bibfnamefont{D.~J.} \bibnamefont{Lacks}},
  \bibinfo{journal}{Phys. Rev. Lett.} \textbf{\bibinfo{volume}{87}},
  \bibinfo{pages}{225502} (\bibinfo{year}{2001}).

\bibitem[{\citenamefont{Maloney and
  Lema\^{i}tre}(2004)}]{Maloney/Lemaitre}
\bibinfo{author}{\bibfnamefont{C.}~\bibnamefont{Maloney}} \bibnamefont{and}
  \bibinfo{author}{\bibfnamefont{A.}~\bibnamefont{Lema\^{i}tre}},
  \bibinfo{journal}{Phys. Rev. Lett.} \textbf{\bibinfo{volume}{93}},
  \bibinfo{pages}{016001} (\bibinfo{year}{2004}),
\bibinfo{author}{\bibfnamefont{C.~E.} \bibnamefont{Maloney}} \bibnamefont{and}
  \bibinfo{author}{\bibfnamefont{A.}~\bibnamefont{Lema\^{i}tre}},
  \bibinfo{journal}{Phys. Rev. E} \textbf{\bibinfo{volume}{74}},
  \bibinfo{pages}{016118} (\bibinfo{year}{2006}).

\bibitem[{\citenamefont{Demkowicz and
  Argon}(2005{\natexlab{a}})}]{Demkowicz/Argon}
\bibinfo{author}{\bibfnamefont{M.~J.} \bibnamefont{Demkowicz}}
  \bibnamefont{and} \bibinfo{author}{\bibfnamefont{A.~S.} \bibnamefont{Argon}},
  \bibinfo{journal}{Phys. Rev. B} \textbf{\bibinfo{volume}{72}},
  \bibinfo{pages}{245205} (\bibinfo{year}{2005}{\natexlab{a}}),
\bibinfo{author}{\bibfnamefont{M.~J.} \bibnamefont{Demkowicz}}
  \bibnamefont{and} \bibinfo{author}{\bibfnamefont{A.~S.} \bibnamefont{Argon}},
  \bibinfo{journal}{Phys. Rev. B} \textbf{\bibinfo{volume}{72}},
  \bibinfo{pages}{245206} (\bibinfo{year}{2005}{\natexlab{b}}).

\bibitem[{\citenamefont{Weiss and Marsan}(2003)}]{Weiss/Marsan:2003}
\bibinfo{author}{\bibfnamefont{J.}~\bibnamefont{Weiss}} \bibnamefont{and}
  \bibinfo{author}{\bibfnamefont{D.}~\bibnamefont{Marsan}},
  \bibinfo{journal}{Science} \textbf{\bibinfo{volume}{299}},
  \bibinfo{pages}{89} (\bibinfo{year}{2003}).

\bibitem[{\citenamefont{Sethna et~al.}(2001)\citenamefont{Sethna, Dahmen, and
  Myers}}]{Sethna/Dahmen/Myers:2001}
\bibinfo{author}{\bibfnamefont{J.~P.} \bibnamefont{Sethna}},
  \bibinfo{author}{\bibfnamefont{K.~A.} \bibnamefont{Dahmen}},
  \bibnamefont{and} \bibinfo{author}{\bibfnamefont{C.~R.} \bibnamefont{Myers}},
  \bibinfo{journal}{Nature} \textbf{\bibinfo{volume}{410}},
  \bibinfo{pages}{242} (\bibinfo{year}{2001}).

\bibitem[{\citenamefont{Sommer et~al.}(1980)\citenamefont{Sommer, Bucher, and
  Predal}}]{Sommer/Bucher/Predal:1980}
\bibinfo{author}{\bibfnamefont{F.}~\bibnamefont{Sommer}},
  \bibinfo{author}{\bibfnamefont{G.}~\bibnamefont{Bucher}}, \bibnamefont{and}
  \bibinfo{author}{\bibfnamefont{B.}~\bibnamefont{Predal}},
  \bibinfo{journal}{J. Phys. Colloque C8} \textbf{\bibinfo{volume}{41}},
  \bibinfo{pages}{563} (\bibinfo{year}{1980}).

\bibitem[{\citenamefont{Inoue et~al.}(1991)\citenamefont{Inoue, Kato, Zhang,
  Kim, and Masumoto}}]{Inoue/others:1991}
\bibinfo{author}{\bibfnamefont{A.}~\bibnamefont{Inoue}},
  \bibinfo{author}{\bibfnamefont{A.}~\bibnamefont{Kato}},
  \bibinfo{author}{\bibfnamefont{T.}~\bibnamefont{Zhang}},
  \bibinfo{author}{\bibfnamefont{S.~G.} \bibnamefont{Kim}}, \bibnamefont{and}
  \bibinfo{author}{\bibfnamefont{T.}~\bibnamefont{Masumoto}},
  \bibinfo{journal}{Mater. Trans. JIM} \textbf{\bibinfo{volume}{32}},
  \bibinfo{pages}{609} (\bibinfo{year}{1991}).

\bibitem[{\citenamefont{Jacobsen et~al.}(1996)\citenamefont{Jacobsen, Stoltze,
  and N{\o}rskov}}]{Jacobsen/Stoltze/Norskov:1996}
\bibinfo{author}{\bibfnamefont{K.~W.} \bibnamefont{Jacobsen}},
  \bibinfo{author}{\bibfnamefont{P.}~\bibnamefont{Stoltze}}, \bibnamefont{and}
  \bibinfo{author}{\bibfnamefont{J.~K.} \bibnamefont{N{\o}rskov}},
  \bibinfo{journal}{Surf. Sci.} \textbf{\bibinfo{volume}{366}},
  \bibinfo{pages}{394} (\bibinfo{year}{1996}).

\bibitem[{\citenamefont{Bailey et~al.}(2004)\citenamefont{Bailey, Schi{\o}tz,
  and Jacobsen}}]{Bailey/Schiotz/Jacobsen:2004a}
\bibinfo{author}{\bibfnamefont{N.~P.} \bibnamefont{Bailey}},
  \bibinfo{author}{\bibfnamefont{J.}~\bibnamefont{Schi{\o}tz}},
  \bibnamefont{and} \bibinfo{author}{\bibfnamefont{K.~W.}
  \bibnamefont{Jacobsen}}, \bibinfo{journal}{Phys. Rev. B}
  \textbf{\bibinfo{volume}{69}}, \bibinfo{pages}{144205}
  (\bibinfo{year}{2004}).

\bibitem[{\citenamefont{Vollmayr-Lee and
  Zippelius}(2005)}]{VollmayrLee/Zippelius:2005}
\bibinfo{author}{\bibfnamefont{K.}~\bibnamefont{Vollmayr-Lee}}
  \bibnamefont{and}
  \bibinfo{author}{\bibfnamefont{A.}~\bibnamefont{Zippelius}},
  \bibinfo{journal}{Phys. Rev. E} \textbf{\bibinfo{volume}{72}},
  \bibinfo{pages}{041507} (\bibinfo{year}{2005}).

\end{thebibliography}

\end{document}